\documentclass[prd,twocolumn,superscriptaddress,floatfix,amsmath,amssymb,floatfix]{revtex4-1}

\usepackage{ifxetex}
\ifxetex
   \usepackage{polyglossia}
   \setdefaultlanguage{english}
   \usepackage{fontspec}
\else
   \usepackage[utf8]{inputenc}
\fi

\usepackage[normalem]{ulem}
\usepackage[english]{babel}
\usepackage{graphicx}
\usepackage{dcolumn}
\usepackage{bm}
\usepackage{verbatim}
\usepackage{mathrsfs}
\usepackage{cancel}
\usepackage{epstopdf}
\usepackage{mathtools}
\usepackage{color}
\usepackage{tensor}
\usepackage{hyperref}

\newcommand{\noprint}[1]{}

\newcommand{\tr}{{\text{tr}}}

\newcommand{\ii}{\mathrm{i}}
\renewcommand{\d}{\mathrm{d}}
\renewcommand{\Re}{\mathrm{Re}}
\renewcommand{\Im}{\mathrm{Im}}
\newcommand{\nn}{\nonumber}

\usepackage{braket}
\usepackage{bbm}

\providecommand{\abs}[1]{\lvert#1\rvert}

\begin{document}

\title{Direct measurement of the two-point function in quantum fields}

\author{José de Ramón}
\affiliation{Institute for Quantum Computing, University of Waterloo, Waterloo, Ontario, N2L 3G1, Canada}
\affiliation{Department of Applied Mathematics, University of Waterloo, Waterloo, Ontario, N2L 3G1, Canada}

\author{Luis J. Garay}
\affiliation{Departamento de F\'isica Te\'orica, Universidad Complutense 
de Madrid, 28040 Madrid, Spain}
\affiliation{Instituto de Estructura de la Materia (IEM-CSIC), Serrano 121, 28006 Madrid, Spain}

\author{Eduardo Mart\'in-Mart\'inez}
\affiliation{Institute for Quantum Computing, University of Waterloo, Waterloo, Ontario, N2L 3G1, Canada}
\affiliation{Department of Applied Mathematics, University of Waterloo, Waterloo, Ontario, N2L 3G1, Canada}
\affiliation{Perimeter Institute for Theoretical Physics, 31 Caroline St N, Waterloo, Ontario, N2L 2Y5, Canada}

\begin{abstract}

We analyze fast interaction cycles in bipartite quantum systems, showing that the statistics in one of the parties (the detector) can be used to determine the two-point correlator of the observable which mediates the coupling in the other (the target). 
We apply the results to the response of particle detectors coupled to quantum fields and subject to this kind of interactions. We show that, in principle, such a setup can be used to experimentally obtain a  direct  evaluation of the Wightman function (both its real and imaginary part) for any field state.

\end{abstract}

\maketitle

\section{Introduction}
 
Particle detectors are localized and controllable quantum systems that are coupled in space and time to quantum fields \cite{Martin-Martinez2018}. Therefore, particle detectors provide us with a framework for the analysis of local aspects in quantum field theory. 
 
Initially formulated to study phenomena such as the well-known Unruh \cite{Unruh:1976aa} and Hawking \cite{Hawking1975} effects, particle detector models have been used in more general questions. Among these  we find the introduction of switching processes, which makes particle detector also suitable to  model the physics of superconducting circuits and the light-matter interaction  (See, e.g., \cite{Martin-Martinez2018}).

Given an interaction Hamiltonian $\hat H$ between the detector and the field, we can define $\hat H'=\chi(t)\hat H$, where $\chi(t)$ is a smooth, integrable function of time that accounts for the physics of switching the detector on and off. These switching functions have played a role in the study of thermalization of particle detectors when smooth and long-time processes are applied \cite{fewsterjorma,Antiunruh,louko2006often}. Conversely, the use of fast and intense couplings, described as ``delta'' interactions,  was introduced for first time in \cite{Hotta:2008aa}, and later used to analyze non-perturbative and perturbative phenomena in RQI-scenarios \cite{PozasJorma,Petarpetandolo,Petarpetandolo1}. As a matter of fact, delta interactions have become one of the few tools that allow us to analyze physics beyond the perturbative regime in quantum field theory (for a summary of non-perturbative approaches see e.g., \cite{GrandBei,Hotta:2008aa,Beilok,Fuentas,Eric}).

Series of fast and intense interactions have been used in the past for operational purposes in relativistic quantum information \cite{Petarpetandolo3}. However, little work has been done  in the analysis of deviations from statistical independence when non-overlapping sequences of interactions are considered.  In a statistical independent series of experiments, such as tossing a coin, the collection of individual probability distributions for each experiment completely determines the total probability distribution of the series. However, if a detector is subjected to two identical interactions with the field at different times, the transition probability of the detector after the whole process cannot be calculated from the excitation probability of each interaction separately. 

This point is not as trivial as it might seem at first sight. The fact that those correlations exists could be argued to be originated from the field change after the first interaction and its backreaction on the detector. However, this hand-wavy argument cannot tell the whole story. Indeed, we will see that the statistical dependence emerges already at leading order in perturbation theory. This means that it cannot come from the state of the field backreacting onto the detector (which happens at subleading orders).

In this paper we start with a very general study where we analyze the behavior of a bipartite system, a detector and a target. We will analyze a situation in which, starting from an uncorrelated state $\hat\rho_{\textsc{d}}\otimes\hat\rho_{\textsc{t}}$, both parties are interacting through evenly distributed fast cycles at leading order in perturbation theory. In section \ref{setup} we show that the deviations from the statistical independence are related with the correlations in the initial state. Concretely, if the detector and the 
target interact linearly by means of an operator acting over the target, say $\hat{O}$,  then the detector's deviations from statistical independence after the interactions are related with the two-point correlator $\braket{\hat{O}(\tau)\hat{O}(\tau')}_{{\hat\rho}_{\textsc{t}}}$.

We show that, provided that the detector's probability distributions after each interaction separately are known, we can directly extract information from the two-point correlator of the operator $\hat O$ in the state $\hat\rho_{\textsc{t}}$.
Further, we show in section \ref{strongfast} that if the interactions are fast enough, then they are completely determined by the correlator evaluated at the times the interactions take place. We finish the general discussion in section \ref{sec:probing}, where we propose a procedure for evaluating the correlator at two different times directly.

 Section \ref{QFT} is devoted to showing that the results from previous sections can be applied in the context of particle detectors. We make use of the so-called Unruh-Dewitt (UDW) model, that describes a detector following a trajectory in space-time and coupled linearly with a quantum field, showing that deviations from statistical independence can also be used to compute the two-point correlator, also known as Wightman function, between the two points at which the interaction takes place. We discuss how the limit of pointlike interactions in space and time can be made sense of, and how much information can be extracted from the quantum field with this set-up, and we list its advantages and limitations respect to the most common set-ups in the particle detector's literature. We conclude and summarize in section \ref{conclusions}.

\section{Set-up}\label{setup}
\subsection{Hamiltonian of the model and description of the system}

We are going to analyze the process of measurement of an observable of a quantum system (target) by coupling another quantum system that acts as a detector.  
In this context, the detector-system bipartite states are density matrices acting on a tensor product of the two Hilbert spaces, $\mathcal{H}=\mathcal{H}_\textsc{d}\otimes\mathcal{H}_\textsc{t}$. 

Let us assume that we couple a degree of the  detector $\hat m$ to an observable of the target $ \hat O$ through a  simple product linear coupling, which in the interaction picture is given by
\begin{equation}\label{hamiltonian}
   \hat H_I=\lambda\chi(\tau)\hat{m}(\tau)\hat{O}(\tau),
\end{equation}
where $\lambda$ is the coupling strength, $\hat{O}$ is an observable over the subspace $\mathcal{H}_\textsc{t}$, $\hat{m}$ is an observable over the subspace $\mathcal{H}_\textsc{d}$ and $\tau$ is a time parameter. The time dependence comes from the fact that we assume that both the detector and the target systems have non-trivial free evolution and thus the operators in the interaction picture are time dependent.

We will assume that the detector degree of freedom $\hat m$ has a discrete spectrum, and, for now, that the operator $\hat O$ for the target system degree of freedom being measured is bounded. We will relax this restriction in further sections when we talk about quantum fields being the target systems and atoms being the detector systems, but we will keep it for the following sections.

Finally $\chi(\tau)$ is a switching function controlling the timescale of the interaction.

In this paper we are interested in computing the excitation probability of the detector subsystem D from its ground state $\ket{g}$ to any other state in $\mathcal{H}_\textsc{d}$. 

Let us consider that the initial state of the detector is the ground state and the bath is in an arbitrary state:
\begin{equation}\label{productstate}
   \hat \rho= \ket{g}\!\bra{g}\otimes \hat{\rho}_{\textsc{t}}.
\end{equation}
From this initial state, we can calculate the transition probability to any excited state at leading order in perturbation theory after the interaction dictated by \eqref{hamiltonian} starting from the following expression:
\begin{align}\label{probprim}
      P^+=&\braket{\hat{U}^{\dagger}(\openone_\textsc{d}-\ket{g}\!\bra{g})\otimes \openone_\textsc{t}\hat U}_{\hat\rho}
\end{align}
where $\hat U=\openone+\hat U^{(1)}+\hat U^{(2)}+\dots$ is the unitary evolution generated by \eqref{hamiltonian} and $\hat U^{(k)}$ is the $k$-term of the Dyson expansion of $\hat U$ in the coupling strength $\lambda$, which is defined as
\begin{align}
  \hat  U^{(k)}=\frac{(-\ii)^k}{k!}\int_{-\infty}^{\infty}...\int_{-\infty}^{\infty} \d\tau_k...\d\tau_1\mathcal{T} \hat H_{I}(\tau_k)...\hat H_{I}(\tau_1).
\end{align}
Here $\mathcal{T}$ indicates the time-ordering operator. We note that the duration of the interaction is encoded  in the switching function $\chi$.

For the initial state \eqref{productstate}, one can simplify \eqref{probprim} taking into account that, e.g.,
\begin{align}
   \nn &\braket{(\openone_\textsc{d}-\ket{g}\!\bra{g})\otimes \openone_\textsc{t}}_{\hat\rho}\\
   &\nn=\tr\left[((\openone_\textsc{d}-\ket{g}\!\bra{g})\otimes \openone_\textsc{t})(\ket{g}\!\bra{g}\otimes \hat{\rho}_{\textsc{t}})\right]\\
    &=\tr\left[(\ket{g}\!\bra{g}-\ket{g}\!\bra{g})\otimes \hat{\rho}_{\textsc{t}}\right]=0.
\end{align}
Cancellations of this kind allow us to  write \eqref{probprim} at leading order
\begin{align}\label{Probsin2}
     \nn P^+={}&\lambda^2 \braket{\hat{U}^{(1)\dagger}(\openone_\textsc{d}-\ket{g}\!\bra{g})\otimes \openone_\textsc{t}\hat U^{(1)}}_{\hat\rho}+\mathcal{O}(\lambda^3)\\
    \nn ={}&\lambda^2\int^{\infty}_{-\infty}\!\!\!\! \d\tau \!\int^{\infty}_{-\infty}\!\!\!\!\!  \d\tau'\chi(\tau)\chi(\tau')  w^c_{g}(\tau,\tau')W_{{\hat{\rho}_{\textsc{t}}}}(\tau,\tau')\\
    &{}+\mathcal{O}(\lambda^3),
\end{align}
where $W_{\hat{\rho}_{\textsc{t}}}(\tau,\tau')$ is the two-point correlator of the observable $\hat{O}$ in the initial state ${\hat{\rho}_{\textsc{t}}}$. It is defined as
\begin{equation}\label{Wiaghtmandef}
    W_{{\hat{\rho}_{\textsc{t}}}}(\tau,\tau')\coloneqq \braket{\hat{O}(\tau)\hat{O}(\tau')}_{{\hat{\rho}_{\textsc{t}}}}.
\end{equation}
On the other hand $w^c_{g}(\tau,\tau')$ is the connected two-point correlator  of the observable $\hat{m}$ in the ground state $\ket{g}$, that is 
\begin{align}
    w^c_{g}(\tau,\tau')\coloneqq \braket{g|\hat{m}(\tau)\hat{m}(\tau')|g}-\braket{g|\hat{m}(\tau)|g}\braket{g|\hat{m}(\tau')|g}.
\end{align}
To simplify the scenario let us model the detector as a two level quantum system and we take $\hat{m}$ as a Pauli observable that does not commute with the detector's free Hamiltonian. If we define the operator $\hat \sigma^-$ as the operator satisfying $\hat \sigma^-\ket{g}=0$, $\hat \sigma^+\coloneqq(\hat \sigma^-)^\dagger$, and $\hat m(0)=\hat \sigma_x$, and then $\hat{m}(\tau)=e^{\ii \Omega \tau} \hat\sigma^+ +e^{-\ii \Omega \tau} \hat{\sigma}^-$  where $\Omega$ is the energy gap between the ground and excited states, we see that $w^c_{g}(\tau,\tau')$ acquires a convenient form
\begin{align}\label{connectedcorrelator}
    w^c_{g}(\tau,\tau')=e^{-\ii\Omega (\tau-\tau')}.
\end{align}

Notice that generalizing this to a higher-dimensional detector is trivial. For example, if the detector is a harmonic oscillator and we choose $\hat{m}(0)$ to be its position operator, that is $\hat{m}(\tau)=e^{\ii \Omega \tau} \hat a^\dagger +e^{-\ii \Omega \tau} \hat{a}$ with $\Omega$ the separation between energy levels, $w^c_{g}(\tau,\tau')$ also has the form of a complex exponential as in \eqref{connectedcorrelator}.

With these simple systems expression \eqref{Probsin2} takes the form of a Fourier transform. From now on we will focus on the case of detectors modelled by two-level systems for simplicity. We define the following  bi-functional over (sufficiently well-behaved \cite{fewsterjorma}) positive switching functions:
\begin{equation}\label{functional}
    \mathcal{W}^\Omega_{{\hat{\rho}_{\textsc{t}}}}[f,g]=\int^{\infty}_{-\infty}\!\!\!\!\! \d\tau \!\int^{\infty}_{-\infty}\!\!\!\!\!  \d\tau'f(\tau)[g(\tau')]^\ast  e^{-\ii\Omega (\tau-\tau')}W_{{\hat{\rho}_{\textsc{t}}}}(\tau,\tau'),
\end{equation}
where the symbol $\ast$ denotes complex conjugation. With this notation at hand the excitation probability takes a more compact form:  
\begin{equation}\label{Probex}
    P^+=\lambda^2\mathcal{W}^\Omega_{{\hat{\rho}_{\textsc{t}}}}[\chi,\chi].
\end{equation}
The functional $\mathcal{W}^\Omega_{{\hat{\rho}_{\textsc{t}}}}[f,g]$ satisfies that:
\begin{align}
    &\mathcal{W}^\Omega_{{\hat{\rho}_{\textsc{t}}}}[f,f]\ge 0,\label{positivewightman}\\
    &\mathcal{W}^\Omega_{{\hat{\rho}_{\textsc{t}}}}[g,f]=\left(\mathcal{W}^{\Omega}_{{\hat{\rho}_{\textsc{t}}}}[f,g]\right)^*\label{conjugatewightman}.
\end{align}
The proofs can be found in appendix \ref{ApA}.

\subsection{Combs of interactions}\label{subsec:combs}

For our purposes, we are going to particularize the switching function $\chi(\tau)$ to a repeated succession of interaction bursts. We will consider a switching function composed of a sequence of $N$ repetitions of a particular time dependent function $\xi(\tau)$. Specifically,
\begin{equation} \label{sumteeth}
     \chi(\tau)=\sum_{l=0}^{N}\xi^{\tau_0}_{l}(\tau).
\end{equation}
Here the notation $\xi^{\tau_0}_{l}(\tau)$ has to be understood as 
\begin{align}
    \xi^{\tau_0}_{l}(\tau)=\xi(\tau-\tau_0-l\zeta),
\end{align}
where $\xi$ is a certain function, which we call ``tooth'' function, $\tau_0$ is the time at which the first tooth is centered, and the rest of the teeth will be centred at $\tau_0+l \zeta$ for $l=1,2,\dots, N$. Hence, $\zeta$ is the constant lapse between each two consecutive interactions.  The excitation probability in these conditions is readily obtained from \eqref{Probex}:
\begin{align}\label{Psum1}
 P^+=&\lambda^2 \sum_{l=0}^{N}\sum_{s=0}^{N}\mathcal{W}^\Omega_{{\hat{\rho}_{\textsc{t}}}}[\xi^{\tau_0}_{l},\xi^{\tau_0}_{s}].
\end{align}
 It is useful to separate this sum in two parts: 
 \begin{enumerate}
     \item The sum of terms where the integrals only involve tooth functions centered at the same time (local terms). These terms are proportional to  $\mathcal{W}[\xi^{\tau_0}_{l},\xi^{\tau_0}_{l}]$, and therefore each one of them independently can be interpreted as the excitation probability of the detector when only one switching process is implemented.
     \item The sum of terms where the integrals involve tooth functions evaluated at different times (non-local terms).
 \end{enumerate}
Now for the non-local terms we see that the sum runs for $l,s$ from $1$ to $N$, with $l\neq s$. Equivalently, we can write this sum with the index $l$ running form $1$ to $N$ and separate the sum for $s<l$ and for with $s>l$. After making this separation explicit, we can write \eqref{Psum1} as
\begin{align}\label{Piposuma}
 \nn&\sum_{\{s\neq l\}}^{N}\mathcal{W}^\Omega_{{\hat{\rho}_{\textsc{t}}}}[\xi^{\tau_0}_{l},\xi^{\tau_0}_{s}]\\
 &=\sum_{l=0}^{N}\sum_{s=l+1}^{N}\mathcal{W}^\Omega_{{\hat{\rho}_{\textsc{t}}}}[\xi^{\tau_0}_{l},\xi^{\tau_0}_{s}]+\sum_{l=0}^{N}\sum_{s=0}^{l-1}\mathcal{W}^\Omega_{{\hat{\rho}_{\textsc{t}}}}[\xi^{\tau_0}_{l},\xi^{\tau_0}_{s}].
\end{align}
Here the first sum can be rearranged by changing the name of the indices. We define $n=s$ and $m=l-s$, in such a way that we have
\begin{equation}
    \sum_{l=0}^{N}\sum_{s=l+1}^{N}\mathcal{W}^\Omega_{{\hat{\rho}_{\textsc{t}}}}[\xi^{\tau_0}_{l},\xi^{\tau_0}_{s}]=\sum_{m=1}^{N} \!\!\sum_{n=0}^{N-m}\mathcal{W}^\Omega_{{\hat{\rho}_{\textsc{t}}}}[\xi^{\tau_0}_{n+m},\xi^{\tau_0}_{n}].
\end{equation}
Similarly, we can write the second sum as
\begin{equation}
    \sum_{l=0}^{N}\sum_{s=0}^{l-1}\mathcal{W}^\Omega_{{\hat{\rho}_{\textsc{t}}}}[\xi^{\tau_0}_{l},\xi^{\tau_0}_{s}]=\sum_{m=1}^{N} \!\!\sum_{n=0}^{N-m}\mathcal{W}^\Omega_{{\hat{\rho}_{\textsc{t}}}}[\xi^{\tau_0}_{n},\xi^{\tau_0}_{n+m}].
\end{equation}

Attending to \eqref{conjugatewightman}, it is clear in this form that both sums in the r.h.s. of \eqref{Piposuma} are complex conjugate of each other. Gathering these results together, we write the excitation probability \eqref{Psum1} as
\begin{align}\label{Pgeneral}
 P^+=\sum_{n=0}^{N}P^+_{\xi,n}+2\lambda^2\Re\,\left[ C_{{\hat{\rho}_{\textsc{t}}}}(\Omega,\tau_0)\right].
\end{align}
Here we have notated the local terms as
\begin{align}\label{Pindividual}
    P^+_{\xi,n}\coloneqq&\lambda^2 \mathcal{W}^\Omega_{{\hat{\rho}_{\textsc{t}}}}[\xi^{\tau_0}_{n},\xi^{\tau_0}_{n}]
\end{align}
and the non-local terms are collected as
\begin{align}\label{correlations}
    C_{{\hat{\rho}_{\textsc{t}}}}(\Omega,&\tau_0)\coloneqq\sum_{m=1}^{N} \!\!\sum_{n=0}^{N-m}\mathcal{W}^\Omega_{{\hat{\rho}_{\textsc{t}}}}[\xi^{\tau_0}_{n+m},\xi^{\tau_0}_{n}].
\end{align}

Note that if instead of a succession of $N$ teeth the switching function were composed of a single tooth $\xi$, the transition probability would be given by just $P^+_{\xi,0}$. In this form it can be readily seen that the local contribution to the total excitation probability consists of the sum of the probability of excitation switched on only for the duration of every individual tooth.

On the other hand $ C_{{\hat{\rho}_{\textsc{t}}}}(\Omega,\tau_0)$ gathers the contributions that comes from the existence of time-correlations in the state $\hat{\rho}_{\textsc{t}}$. 

Note that equation \eqref{Pgeneral} works for any general comb interactions even if the the two-point function is not stationary (i.e., even if it does not depend only on the difference of times). We conclude this section with some remarks specific to the situation when we impose stationarity.

\subsection{Stationary correlations}

 We can consider situations in which the two-point function is stationary, that is,
\begin{equation}
    W(\tau,\tau')=W(\tau-\tau').
    \end{equation}
In other words, the correlations between two events depend only on the difference of times between them.
It is interesting to consider the stationarity condition as it is fulfilled in paradigmatic scenarios in physics of detectors interacting with quantum fields, which is our main goal here. Such scenarios comprise, for instance, the interaction of inertial detectors with vacuum and thermal baths, as well as the interaction of uniformly  accelerated detectors with vacuum, as we will discuss in more detail in section \ref{sec:probing}.

 When the situation is stationary it is very easy to see how the information about the target correlations is imprinted in the excitation probability. This allows us to conceive a simple theoretical protocol to extract this information.

 For stationary correlations we can see that neither \eqref{Pindividual} nor the different terms in \eqref{correlations}  depend on $n$ or $\tau_0$.  Indeed,

\begin{align}\label{invariance}
  \nn &  \mathcal{W}^\Omega_{{\hat{\rho}_{\textsc{t}}}}[\xi^{\tau_0}_{n+m},\xi^{\tau_0}_{n}]\\
  \nn &=\int^{\infty}_{\!\!-\infty}\int^{\infty}_{\!\!-\infty}\!\!\!\!\! \d\tau \d\tau' \xi(\tau\!-\!\tau_0\!-(n+m)\zeta)\xi(\tau'\!-\!\tau_0\!-n\zeta)\\
\nn & \qquad\qquad\qquad \times e^{-\ii\Omega (\tau\!-\!\tau')}W_{\hat{\rho}_{\textsc{t}}}(\tau\!-\!\tau') \\
 \nn&  =\int^{\infty}_{\!\!-\infty}\int^{\infty}_{\!\!-\infty}\!\!\!\!\! \d\tau \d\tau' \xi(\tau-m\zeta)\xi(\tau') e^{-\ii\Omega (\tau\!-\!\tau')}W_{\hat{\rho}_{\textsc{t}}}(\tau\!-\!\tau')\\
 & \hfill = \mathcal{W}^\Omega_{{\hat{\rho}_{\textsc{t}}}}[\xi^{0}_{m},\xi^{0}_{0}].
 \end{align}
Using the invariance \eqref{invariance} we can see that \eqref{Pgeneral} simplifies to
\begin{equation}\label{Pstacionary}
    P^+=NP^+_{\xi,0}+2\lambda^2\Re\,\left[ C_{\hat{\rho}_{\textsc{t}}}(\Omega,0)\right],
\end{equation}
where
\begin{equation}
    P^+_{\xi,0}=\lambda^2\mathcal{W}_{{\hat{\rho}_{\textsc{t}}}}[\xi^{0}_{0},\xi^{0}_{0}]
\end{equation}
is again the excitation probability associated to one single interaction and
\begin{align}\label{Cgeneral}
     &\nn C_{\hat{\rho}_{\textsc{t}}}(\Omega,0)=\sum_{m=1}^{N}  \left(N-m\right) \mathcal{W}^\Omega_{{\hat{\rho}_{\textsc{t}}}}[\xi^{0}_{m},\xi^{0}_{0}].
\end{align}
is the time-correlation term. 

In this case, it is easier to understand the nature of the two summands in \eqref{Pstacionary}: if we considered $N$ statistically independent switching experiments with a single tooth, the probability of finding one excitation  after $N$ independent switching experiments follows a binomial distribution, which for small individual transition probabilities can be approximated as follows:
\begin{equation}
    P^+=\binom{N}{1}P^+_\xi(1-P^+_\xi)^{N-1}\approx NP^+_\xi.
\end{equation}
In our case this approximation holds since we are in the perturbative regime hence $P^+_\xi \ll 1$ . 

This observation allows us to identify the first summand in \eqref{Pgeneral} as the probability of finding an excitation in $N$ uncorrelated repetitions of a single switching experiment. Hence,  the second summand corresponds to the deviation from the uncorrelated scenario given by the time correlations in the field when we consider a succession of $N$ switching teeth.

\section{Strong and fast couplings}\label{strongfast}

In this section we will consider that the comb of interactions consists of short and intense teeth. We will see that this kind of interactions allows one to probe the two-point correlator with arbitrary precision.

\subsection{Single tooth interaction}
 We shall calculate the excitation probability when the switching function is short and intense but with a finite width, which we will call $\eta$. More concretely we are interested in the behavior of the excitation probability when this parameter $\eta$ is small. We can parametrize a switching function with those characteristics in a sufficiently general form as:
\begin{equation}\label{nascent}
    \chi=\varphi_\eta=\frac{1}{\eta}\varphi(\tau/\eta),
\end{equation}
where $\varphi$ is an integrable, positive function which is normalized to one. Notice that in order to avoid the introduction of additional timescales we must endow the switching functions with  dimensions of inverse time. The one-parameter characterization \eqref{nascent} leads to families of nascent delta functions. Indeed delta-switching have been used in the literature in perturbative \cite{Hotta:2008aa,Petarpetandolo,PozasJorma} and non-perturbative \cite{Petarpetandolo1} calculations.  If we smear a function $f(\tau)$ with any member of this family and we take the limit $\eta\to0$, we have
\begin{align}
    \nn\lim_{\eta\to 0} \int\frac{\d\tau}{\eta}\varphi(\tau/\eta)f(\tau)=&\lim_{\eta\to 0} \int\d\bar\tau\varphi(\bar\tau)f(\eta\bar\tau)\\
    =&f(0)\int\d\bar\tau\varphi(\bar\tau)=f(0),
\end{align}
where we have performed the change of variables $\bar\tau=\tau/\eta$ and we have assumed that $\varphi(\tau)f(\eta\tau)$ is dominated by an integrable function for all $\eta$, in such a way that we can take the limit inside the integral \cite{bartle2014elements}. 

If we calculate the action of bi-functionals  of the form \eqref{functional} over members of these families we get:
\begin{align}
   \nn \mathcal{W}^\Omega_{{\hat{\rho}_{\textsc{t}}}}[\varphi_\eta,\varphi'_\eta]=&\int^{\infty}_{-\infty}\!\!\!\!\!\d\bar{\tau} \int^{\infty}_{-\infty}\!\!\!\!\!  \d\bar{\tau}'\varphi(\bar{\tau})\varphi'(\bar{\tau}') \\
   &\times e^{-\ii\Omega\eta (\bar{\tau}-\bar{\tau}')}W_{{\hat{\rho}_{\textsc{t}}}}(\eta\bar{\tau},\eta\bar{\tau}'),
\end{align}
but in this case one has to be careful when taking the limit $\eta\to 0$, as $W_{{\hat{\rho}_{\textsc{t}}}}$ is not a function (but a distribution) in general.

\subsection{Comb of interactions}\label{kicks}

We will see that under those assumptions the behavior of the non-local terms and of the individual excitation probabilities in \eqref{Pgeneral} are, in principle, qualitatively different as the latter can diverge in relativistic scenarios. Also the non-local terms are well defined in the limit $\eta\to 0$ under some assumptions and, as we argue in section \ref{sec:probing}, can be used for probing the two-point function.  We leave the study of divergent two-point distributions for appendix \ref{singlekick}, where we pay attention to the behavior of the individual excitation probabilities and their relation with the limit $\eta\to 0$ in the context of quantum field theory.

In this subsection we apply a comb of interactions of the form \eqref{sumteeth}, where this time the teeth are nascent delta functions of the form \eqref{nascent}. We restrict our study  here to the time-correlations involving the non-local terms given in equation \eqref{correlations}.
Then the total switching function acquires the form:
\begin{equation} 
     \chi(\tau)=\sum_{l=0}^{N}\varphi^{\tau_0}_{l,\eta}(\tau),
\end{equation}
where 
\begin{align}
    \varphi^{\tau_0}_{l,\eta}(\tau)=\frac{1}{\eta}\varphi\left(\frac{\tau-l\zeta-\tau_0}{\eta}\right).
\end{align}

Substituting in \eqref{correlations} we get
\begin{align}
    C_{{\hat{\rho}_{\textsc{t}}}}(\Omega,&\tau_0)\coloneqq\sum_{m=1}^{N} \!\!\sum_{n=0}^{N-m}\mathcal{W}^\Omega_{{\hat{\rho}_{\textsc{t}}}}[\varphi^{\tau_0}_{n+m,\eta},\varphi^{\tau_0}_{n,\eta}].
\end{align}

 The expression of the action of the functionals $\mathcal{W}^\Omega_{{\hat{\rho}_{\textsc{t}}}}$ over the nascent delta switchings within the non-local terms is
 \begin{align}\label{uvedoble}
\nn&\mathcal{W}^\Omega_{{\hat{\rho}_{\textsc{t}}}}[\varphi^{\tau_0}_{n+m,\eta},\varphi^{\tau_0}_{n,\eta}]\\
 \nn  &= e^{-\ii\Omega\zeta m} \int^{\infty}_{-\infty}\int^{\infty}_{-\infty}\!\!\!\!\! \d\tau \d\tau' \varphi(\bar\tau)\varphi(\bar\tau')  e^{-\ii\Omega\eta (\tau-\tau')}\\
    &\quad\times W_{\hat{\rho}_{\textsc{t}}}\left(\eta\bar\tau+\tau_0+(n+m)\zeta,\eta\bar\tau'+\tau_0+n\zeta\right),
\end{align}
where we have performed the following changes of the variables:
\begin{align}
    \bar\tau&=\frac{\tau-(n+m)\zeta-\tau_0}{\eta},\quad\bar\tau'=\frac{\tau'-n\zeta-\tau_0}{\eta}.
\end{align}

There is a sufficient condition for \eqref{uvedoble} to be well defined for any $\eta\neq 0$. Namely, if given some $r>0$ there exists some $\eta_0$ such that if $\eta>\eta_0$, then
\begin{align}\label{condition}
   \abs{ W_{\hat{\rho}_{\textsc{t}}}(\eta\bar\tau,\eta\bar\tau'+r)}\leq C
\end{align}
for some $\bar{\tau},\bar{\tau}'$, then the two-point correlation function does not diverge for any $r>0$, and more concretely the integral in \eqref{uvedoble} is dominated absolutely, as it is well defined for any $\eta\neq 0$. Note that condition \eqref{condition} is automatically satisfied if $\hat{O}$ is a bounded operator  \cite{berberian1974lectures}.

If we consider the case $r=0$, then assuming the condition \eqref{condition} excludes a lot of interesting cases. As a paradigmatic example, in Quantum Field theory the Wightman function (the  two-point field correlator) can be written as the sum of a regular part and a singular contribution (which is a distribution that in general will not satisfy \eqref{condition} \cite{wald1994quantum}). However, the integral \eqref{uvedoble} will still be well defined for sufficiently regular switching functions. One may wonder if it may be tricky to consider in that case the limit of a delta switching and if such a limit may commute with the integral in \eqref{uvedoble}. However, even in the case that \eqref{condition} is not satisfied, one can make sense of the limit commutation considering that the delta limit arises from the pointlike limit of a sequence of switching functions of constant area that is symetrically made widthless. A discussion along these lines can be found in \cite{pozas2016entanglement}. For sufficiently symmetric regularizations of the Dirac deltas  (understood as the limit of symmetric regular switching functions as nascent deltas), the limit $\eta\rightarrow 0$ can be taken under the integral sign also in this case.

In conclusion, be it because the condition \eqref{condition} is fulfilled or because we understand the nascent deltas as symmetric limits of smooth enough switching functions, we can take the limit $\eta\to 0$ inside the integral sign, and write
\begin{align}
    \lim_{\eta\to0}\mathcal{W}^\Omega_{{\hat{\rho}_{\textsc{t}}}}[\varphi^{\tau_0}_{n+m,\eta},\varphi^{\tau_0}_{n,\eta}]=W_{\hat{\rho}_{\textsc{t}}}\left(\tau_0+(n+m)\zeta,\tau_0+n\zeta\right).
\end{align}
Therefore we conclude that the non-local contribution to the excitation probability has a well defined behavior in the limit $\eta\to 0$ and we write
\begin{align}\label{non-local}
   &\nn\lim_{\eta\to0} C_{{\hat{\rho}_{\textsc{t}}}}(\Omega,\tau_0)\\
   &=\sum_{m=1}^{N}e^{-\ii\Omega \zeta m} \!\!\sum_{n=0}^{N-m}W_{\hat{\rho}_{\textsc{t}}}\left(\tau_0+(n+m)\zeta,\tau_0+n\zeta\right).
\end{align}
This behavior with $\eta$ is crucial. It means that, for series of fast and intense kicks the deviations from statistic independence in the excitation probability of the detector only depend on the correlations in the initial state $\hat{\rho}_{\textsc{t}}$ between the events when they take place. 

This result, that seems trivial, is what will allow to reconstruct the Wightman function from the excitation probability of the detector, as we will see in section \ref{sec:probing}.

To conclude this subsection we remark that, for simplicity in the expressions, we have applied evenly distributed interactions in time. Nonetheless, it is easy to check that the former statement remains true even when the interactions are not evenly distributed.

\section{Probing the two-point correlator with fast kicks}\label{sec:probing}
In this section we finally propose a method for probing the two-point function of the state $\hat{\rho}_{\textsc{t}}$. The method, as it has been advanced in previous sections, will consist in predicting the value of the two-point correlator between two events  by kicking the system T sharply and intensely with a nascent delta switching centered in those two events and measuring their excitation probabilities.

A first characteristic of this method is that we have to be able to measure the excitation probability with single kicks, and measure the excitation probability with two kicks for the same couple of events.  This requires to be able to ensure that if we repeat the experiment the two events, located at  $\tau_1,\tau_2$, are equivalent to other two events located at  $(\tau_1+\alpha,\tau_2+\alpha)$, where $\alpha$ is an arbitrary lapse of proper time.

Besides, the accurate measurement of an excitation probability requires a huge amount the experiments. So if we want to predict the two-point correlator we need to repeat the experiment as many times as necessary and be sure that the quantities that we are measuring do not change if we perform the experiment at two different times.

All these considerations lead to the necessity of considering almost exclusively stationary situations, as it was pointed out in section \ref{subsec:combs}.

\subsection{Probing of the two-point correlator}
The theoretical protocol that can be used, in principle, for probing the two-point correlator by applying two fast kicks requires the following assumptions:
\begin{itemize}
    \item We can measure the excitation probability of the detector regardless of the number of interactions we are applying. More concretely, this means that we can measure the excitation probability when both a double-kick and a single-kick interaction are implemented at any time of our election. In our notation, we have access to the quantities $P^+_{\xi,0}$, $P^+_{\xi,1}$ and $P^+$. The indexes $0,1$ denote that we are applying the same interaction, described by the tooth function $\xi$, at two different times separated by a lapse $\zeta$.
    \item We are free to change the parameters of the detector: the energy gap $\Omega$ and the lapse between interactions $\zeta$. 
    \item Besides, we assume that we can shrink the tooth functions in the way we studied in section \ref{strongfast}, that is, we can choose tooth functions $\xi$ in such a way they act like nascent delta functions, $\varphi_\eta$. 
    \item  We also assume that we can kick the target with intense interactions but not intense enough for coming out from the grounds of perturbation theory in the coupling constant $\lambda$.
\end{itemize}

Under these conditions, we particularize the expression \eqref{sumteeth} for the switching function for $N=2$ and apply our results straightforwardly. First, the excitation probability in \eqref{Pgeneral} takes the form
\begin{align}
   \nn P^+&= P^+_{\xi_1}+P^+_{\xi_0}+2\lambda^2\Re\,\left[ C_{{\hat{\rho}_{\textsc{t}}}}(\Omega,\zeta,2,\tau_0)\right]\\
& = P^+_{\xi_1}+P^+_{\xi_0}+ 2\lambda^2\Re\,\left[ \mathcal{W}^\Omega_{\hat{\rho}_{\textsc{t}}}[\xi^{\tau_0}_{1},\xi^{\tau_0}_{0}]\right] .
\end{align}
 We define now the following function:
\begin{align}\label{estadistico}
 \nn    \mathcal{S}(\Omega,\tau_0):=& {}\frac{1}{2\lambda^2}\left( P^+-P^+_{\xi_1}-P^+_{\xi_0}\right)\\
 =& {}\Re\,\left[ \mathcal{W}^\Omega_{\hat{\rho}_{\textsc{t}}}[\xi^{\tau_0}_{1},\xi^{\tau_0}_{0}]\right],
\end{align}
which only involves measurable quantities as we have claimed. By doing so we realize that we can obtain the value of $\Re\,\left[ \mathcal{W}^\Omega_{\hat{\rho}_{\textsc{t}}}[\xi^{\tau_0}_{1},\xi^{\tau_0}_{0}]\right]$ just measuring the excitation probability in several processes. 

Finally, for tooth functions of the form \eqref{nascent} we calculate the delta-switching limit of \eqref{estadistico}: 
\begin{align}
    \nn  &\lim_{\eta\to0}\Re\,\left[ \mathcal{W}^\Omega_{\hat{\rho}_{\textsc{t}}}[\varphi^{\tau_0}_{1,\eta},\varphi^{\tau_0}_{0,\eta}]\right]
    \\
    &\nn =\Re[e^{-\ii\Omega\zeta}W_{{\hat{\rho}_{\textsc{t}}}}(\zeta+\tau_0,\tau_0)]\\
    &\nn =\cos(\Omega\zeta)\Re[W_{{\hat{\rho}_{\textsc{t}}}}(\zeta+\tau_0,\tau_0)]\\
    &\quad+\sin(\Omega\zeta)\Im[W_{{\hat{\rho}_{\textsc{t}}}}(\zeta+\tau_0,\tau_0)].
\end{align}
In these conditions we can build the whole two-point correlator by forcing the quantity $\Omega\zeta$ to take some concrete values, that is, if we synchronize the detector with the lapse between interactions. On the one hand, if $\Omega\zeta=2\pi k$ where $k$ is an integer, then we have that
\begin{align}\label{Real}
   \lim_{\eta\to0} \mathcal{S}(2\pi k\zeta^{-1},\tau_0)=\Re[W_{{\hat{\rho}_{\textsc{t}}}}(\zeta+\tau_0,\tau_0)].
\end{align}
On the other hand, if we synchronize the detector by choosing the energy gap to fulfill $\Omega\zeta=2\pi k'+\pi/2$, with $k'$ a different integer, we have that
\begin{align}\label{Imaginary}
   \lim_{\eta\to0}  \mathcal{S}\left((2\pi k'\!\!+\pi/2)\zeta^{-1},\tau_0\right)=\Im[W_{{\hat{\rho}_{\textsc{t}}}}(\zeta+\tau_0,\tau_0)].
\end{align}
Thus we can build the whole two-point correlator in terms of measurable quantities:
\begin{align}
    \nn &W_{{\hat{\rho}_{\textsc{t}}}}(\zeta+\tau_0,\tau_0)\\
    &=\lim_{\eta\to0} \left[\mathcal{S}(2\pi k\zeta^{-1},\tau_0)+\ii\mathcal{S}\left((2\pi k'\!\!+\pi/2)\zeta^{-1},\tau_0\right)\right].
\end{align}
In sight of this result, we see that we can approximately measure the two-point correlator between whatever two events by measuring the excitation probability of each interaction separately, and the excitation probability of the joint interaction. Further, as we pointed out at the beginning of this section the expression is completely meaningful. In terms of stationary correlations the statistic $\mathcal{S}$ takes the form
\begin{align}
    \mathcal{S}(\Omega,0)=\frac{1}{2\lambda^2}\left( P^+-2P^+_{\xi_1}\right).
\end{align}
So, once we know $P^+_{\xi_1}$, which can be measured in an independent experiment, we have full information about the two point function by measuring the excitation probability $P^+$ with the comb of nascent deltas:
\begin{align}\label{probing}
     & W_{{\hat{\rho}_{\textsc{t}}}}(\zeta)=\lim_{\eta\to0} \left[\mathcal{S}(2\pi k\zeta^{-1},0)+\ii\mathcal{S}\left((2\pi k'\!\!+\pi/2)\zeta^{-1},0\right)\right].
\end{align}

\section{Detector probes in quantum field theory }\label{QFT}

The formalism developed in previous sections has applications in Quantum Field Theory in the context of particle detector models. Among these, the most notorious is the so-called Unruh-DeWitt model \cite{Unruh:1976aa,dewitt1979quantum}, which describes the interaction of a two-level system coupled linearly to a scalar quantum field. In this section we are interested in how much information can be extracted from a quantum field with fast kicks. 

We will see, in the context of the UDW model, that we can evaluate the field two-point correlator (the so-called Wightman function) between two events that are connected by the trajectory of a detector $\mathsf{x}(\tau)=(t(\tau),\bm{x}(\tau))$, namely,
\begin{align}
    W(\tau)=\braket{\hat{\phi}\left(\mathsf x(\tau)\right)\hat{\phi}\left(\mathsf x(0)\right)}_{{\hat\rho_{\textsc{t}}}}.
\end{align}.

The  Unruh-DeWitt detector is a localized two-level system that couples locally to a scalar field. While simple, this model captures the fundamental features of the light-matter interaction when exchange of angular momentum is not relevant \cite{pozas2016entanglement,Martin-Martinez2018}. 

$\mathcal{H}_{\textsc{t}}$ is now the Hilbert space that describes the state of the scalar quantum field.
In the original model the detector is a point-like two-level system that follows a certain time-like trajectory in space-time and interacts with the quantum field. The Hamiltonian that describes such a situation is given by the form \eqref{hamiltonian}, where the operator $\hat{O}$ that mediates the interaction is simply the amplitude of the field evaluated along the trajectory:
\begin{align}
    \hat{O}(\tau)=\hat{\phi}(\mathsf{x}(\tau)).
\end{align}

The case classically most often studied in the literature is the case of a  detector that interacts for a long time with the field. The two-point correlator then is often assumed to be stationary so that the detector  reaches the equilibrium with the field in the adiabatic limit. For stationary correlations, it can be shown \cite{fewsterjorma} that the asymptotic behavior of the excitation probability for long times (and at leading order in the coupling constant $\lambda$) takes the form
\begin{align}\label{usualP}
    P^+\sim (\text{Interaction time})\times \lambda^2\tilde{W}(\Omega).
\end{align}
Here $\tilde{W}(\Omega)$ is the Fourier transform of the two-point correlator evaluated at the energy gap of the detector. 

Note that a similar calculation for the deexcitation probability leads to
\begin{align}
    P^-\sim (\text{Interaction time})\times \lambda^2\tilde{W}(-\Omega).
\end{align}
The distributional nature of $W$ in the case of quantum fields requires a careful approach to understanding expression \eqref{usualP}. However, this expression can be made fully meaningful in most common situations by understanding the process as a limit of long-time, smooth switching processes (adiabatic limit) \cite{fewsterjorma}. However, some subtleties related with this issue will appear in the context of our formalism, and will be discussed later in this section.

One may wonder  how much information about the field is encoded in \eqref{usualP}, in other words, how much information about the field does the detector extract from a stationary field state in the adiabatic limit. First we observe that the quantity $C(\Omega)=\tilde{W}(\Omega)-\tilde{W}(-\Omega)$, the anti-symmetric part of the Fourier transform of the Wightman distribution, has the following expression:
\begin{align}
   C(\Omega)=\int_{\infty}^\infty\d\tau e^{-\ii\Omega\tau}[\hat{\phi}\left(\mathsf x(\tau)\right),\hat{\phi}\left(\mathsf x(0)\right)],
\end{align}
that is, the Fourier transform of the commutator of the fields. For fields fulfilling canonical commutation relations  this quantity is a c-number (a multiple of the identity). Thus it is independent of the state of the field.  Knowing $C(\Omega)$ allows one to probe the classical features of the quantum field. Among these, we can consider, for instance the mass of the field and the speed of propagation. The field commutator of the fields is a solution of the homogeneous equation of motion \cite{wald1994quantum,birrell1984quantum,takagi1986}, thus containing classical information.

In contrast, the symmetric part of the transformed Wightman function $S(\Omega)=\tilde{W}(\Omega)+\tilde{W}(-\Omega)$ contains information about the state of the field. In previous literature, it has been common to consider the interaction of field states with detectors following trajectories that  are associated with symmetries of the space-time, namely, accelerated detectors (related with boosts), inertial trajectories  (related with space-time translations), and  circular motion (related with rotations). Particular efforts were put on probing the thermal character of accelerated detectors interacting with the Minkowski vacuum, which is commonly known as Unruh effect \cite{Unruh:1976aa}, as well as the characteristics of the radiation emitted by black holes \cite{Hawking1975}. 

 All these features can be analyzed from equilibration processes corresponding to equation \eqref{usualP}. However, there are many other interesting features that would be missing in such analyses.  Waiting for equilibration formally requires  to take the long-time interaction limit, which makes the nature of the predictions non local. The response of the detector will omit the details of the local structure of the field. Some questions, as how different effects dominate at different time scales cannot be addressed with this setup. For instance, an accelerated detector interacting with a broad class of states will finish in a thermal state at the Unruh temperature \cite{bievre2006}. To a large extent, everlasting accelerated detectors can erase any non-Lorentz invariant effects in the theory, as shown in \cite{Carballo-Rubio:2018aa}. In addition, this non-locality implies that the detector is not obtaining information of the Wightman function at some given space-time point. Instead the whole world line of the detector is integrated and the response of the detector becomes highly dependent on the particular trajectory of the detector, therefore yielding different predictions for different trajectories when the detector crosses the same space-time points. Whereas this fact is an advantage for the analysis of the Unruh effect, as an analogy between moving detectors and alternative quantization schemes can be established, it poses a challenge to direct measurements of the correlation functions of the field, since for a given detector phenomenology, we cannot separate its dependence on the detector trajectory from the dependence on the field state.

Analyzing the response of the detector for finite time interactions can address the limitations of the analyses mentioned above. The procedure outlined in previous sections allows us to imagine a direct measure  of the correlations of the quantum field between two space-time events, thus avoiding the ambiguity of whether the pull-back on a particular trajectory or the actual Wightman function is being probed. 

Henceforth, we will analyze the application of the formalism developed in previous sections to the UDW model, as well as discuss its advantages and its limitations.

First, taking into account the result \eqref{probing} we realize that we can calculate the value
\begin{align}\label{Wightman2}
     W_{{\hat\rho_{\textsc{t}}}}(\zeta+\tau_0,\tau_0)=\braket{\hat{\phi}\left(\mathsf{x}(\zeta+\tau_0)\right)\hat{\phi}\left(\mathsf{x}(\tau_0)\right)}_{{\hat\rho_{\textsc{t}}}}
\end{align}
as a difference of excitation probabilities. 

This model assumes complete knowledge of the trajectory of the detector $\mathsf{x}(\tau)$, and we assume that the inertia of the detector is larger than the momentum exchange with the field, i.e., we neglect any possible backreaction of the field on the detector's trajectory. 
 
Knowing the trajectory of the detector parametrized in its proper time  $\tau$ allows us to fix the two space-time events we want to probe. In other words, once we know $\zeta$, the proper lapse of time between the two interactions, and the initial time $\tau_0$, we know the exact points $\mathsf{x}(\zeta+\tau_0)$ and  $\mathsf{x}(\zeta+\tau_0)$ between the correlations are being measured.

There are several subtle aspects when we carry out these analyses in the context of quantum field theory:

\begin{enumerate}
    \item At first order in perturbation theory, we can only evaluate the correlations between two spatial regions at two different times. We do not have access to the $n$-point function which in general implies that we have no complete knowledge of the quantum field. Nonetheless, it is well known that in certain situations the $n$-point function can be constructed from the two-point Wightman function. This is the case for the so-called quasi-free states, also known as Gaussian states \cite{wald1994quantum}.
    \item We cannot evaluate the correlation function between  space-like or light-like separated regions, as any physical detector can only follow time-like trajectories. 
\end{enumerate}

Among these, a particularly important technical aspect is that the value predicted in \eqref{Wightman2} cannot be thought as the value of a function overall, but as the outcome of a distribution acting over narrow smearing functions. However, the point-like information  about the Wightman function can be made meaningful when it is understood as a limit of regular finite-smearing distributions. As an example, it is well known that any regularization of the Wightman distribution of the vacuum of a Klein-Gordon field is divergent in the coincidence limit when the regularization is removed \cite{wald1994quantum,birrell1984quantum}.  However this does not present a problem in this case as the non-local terms in \eqref{uvedoble} do not involve the coincidence limit, so the limit of the switching functions to delta distributions is well defined. 

 Importantly,  we recall that \eqref{estadistico} involves the value of the difference of individual excitation probabilities. Both the excitation probability of the combined process and the individual excitation probabilities diverge in the delta-switching limit, but these divergences are spurious and point-independent:  the different excitation probabilities at different times diverge but their difference remains finite. This allows us to, in principle, suggest possible experimental protocols that obtain the value of the asymptotic expression $W_{{\hat\rho_{\textsc{t}}}}(\zeta+\tau_0,\tau_0)$ as $\eta\to0$. In appendix \ref{singlekick} we develop, for completeness, the asymptotic behavior of the individual excitation probabilities in the delta switching limit for a wide variety of cases.

\section{Conclusions}\label{conclusions}

In the context of a quantum system probed by a quantum detector, we have shown that we can use the detector response to time-separated interactions as a tool to extract information from the correlations in the target system by analyzing the statistics in the detector system. 

We have analyzed the general form of the excitation probability of the detector from the ground state to any other state at leading order in perturbation theory, focusing in the simplest case of a two-level system. 

We can write this excitation probability after two fast interactions as the sum of the individual excitation probabilities plus deviations from statistical independence.  These deviations encode crucial information about the target two-point correlation function smeared over the switching function that has two peaks centered at different times. We have further analyzed the particularly relevant case where the correlations are stationary, which corresponds to interesting cases such as thermalization or any other equilibration process. 

Next, we have taken the formal limit when the interactions are fast and intense, thus resembling delta distributions.  In this limit the correlations are only dependent on the two-point correlator evaluated at the times where the interactions take place.

Under the former conditions, we developed a detection protocol in which both the real and imaginary components of the two-point correlator can be directly measured in two different times. This protocol assumes that once the lapse of time between interactions $\zeta$ is fixed, we can change the energy gap of the detector $\Omega$. Both the real (imaginary) part of the correlator can be measured if we synchronize the energy gap of the detector to perform an even (odd) number of free cycles between interactions.

Finally, we have applied these results to particle detectors coupled to quantum fields. Assuming the UDW model, this method allows one to measure the  Wightman function between two events that are time-like separated for a wide variety of states and theories. 

We have discussed the mathematical subtleties of evaluating an object of distributional nature in a point, then seeing that this can be done as far as the Wightman distribution does not diverge inside the light-cone. Of course, this is the case for the most common theories that can be found in the literature.

We have also compared this method of measuring the Wightman distribution with the more conventional way, that for a stationary state, consists in waiting for the detector to reach equilibrium. This fast-interaction approach avoids taking the long-time limit, leading thus to measurements that do not involve the non-local character of equilibration. In summary, finite time interactions can be used to probe the Wightman function of a quantum field. This has the advantage that we are able to measure the correlation function in a space-time localized manner in two points, rather than its pullback over a full trajectory, which is the information one gets when waiting for equilibration.

\acknowledgments

This work has been supported in part by the MINECO (Spain) projects FIS2014-54800-C2-2 and  FIS2017-86497-C2-2 (with FEDER contribution). The work of E.M.-M. is supported by the Natural Sciences and Engineering Research Council of Canada through the Discovery program. E.M.-M. also gratefully acknowledges the funding of his Ontario Early Research Award.
\appendix

\section{Proof of properties of the Wightman functional}\label{ApA}

In this appendix we prove the properties \eqref{positivewightman} and \eqref{conjugatewightman}, that play a key role in the derivation of equation \eqref{probing}.

We are going to prove first that
\begin{equation}
\mathcal{W}_{\rho}[f,f]>0.
\end{equation}

 Let us define the following operator,
\begin{align}
   \hat{O}^\Omega_{f}=\int^{\infty}_{-\infty}\!\!\!\!  \d\tau f(\tau)  e^{-\ii\Omega \tau}\hat{O}(\tau).
\end{align}
In terms of this operator we can write 
\begin{equation}\label{Wphi}
\mathcal{W}_{\rho}[f,g]=\braket{\hat{O}^\Omega_{f}{\hat{O}^\Omega_{g}}},
\end{equation}
we realize that if $f=g$ then $\mathcal{W}_{\rho}[f,f]$ is the average of $\hat{O}^\Omega_{f}{\hat{O}^\Omega_{f}}{}^\dagger$, that is a positive operator.

Next, We show that  
\begin{equation}
\mathcal{W}_{\rho}[g,f]=\mathcal{W}_{\rho}[f,g].
\end{equation}
This property can be readily proved by taking the complex conjugate in \eqref{Wphi},
\begin{equation}
\mathcal{W}_{\rho}[f,g]=\braket{\hat{O}^\Omega_{f}{\hat{O}^\Omega_{g}}{}^\dagger}_{\rho}^*=\braket{\hat{O}^\Omega_{g}{\hat{O}^\Omega_{f}}{}^\dagger}_{\rho}=\mathcal{W}_{\rho}[g,f].
\end{equation}

\section{Transition probability of a single kick} \label{singlekick}

In this appendix we analyze the excitation probability of a detector coupled to a scalar Klein-Gordon field with a switching function of the form \eqref{nascent}. In this context the excitation probability will not be well defined in the limit $\eta\to 0$. This will be caused, as we will see, by the well-known ultra-violet divergent character of relativistic quantum field theories \cite{wald1994quantum}.

 It is clear that if the interaction time goes to zero without varying the intensity of the interaction, then the excitation probability will vanish in the instantaneous interaction limit. However, it is a very different case if we increase the intensity of the the interactions as they get shorter in a delta switching limit. 

 It is well known \cite{schlicht2004considerations,louko2006often} that taking arbitrarily sudden interactions causes UV divergences in the model. Regarding the nature of the relation between the ratio between the intensity of the interaction and the interaction time we have three options. First
as we mentioned above, the interaction time decreases as the intensity increases, but at a slower rate, so we remain in the case where the excitation probability vanishes as the interaction time vanishes.
Second, the intensity increases faster than the interaction time decreases. In this case a divergence in the excitation probability is expected regardless of the UV structure of the theory.
We are interested in the third option: If the interaction intensity increases at the same rate as the interaction time decreases, as in \eqref{nascent}, then the excitation probability has an UV divergence.

Indeed if we use $\chi=\delta_{\tau_0}$ as switching function in the excitation probability \eqref{Probex}, we obtain
\begin{equation}
    P^{+}=\lambda^2\mathcal W_{\hat{\rho}_{\textsc{t}}}(\delta_{\tau_0},\delta_{\tau_0})=\lambda^2W_{\hat{\rho}_{\textsc{t}}}(\tau_0,\tau_0),
\end{equation}
which turns out to be divergent in the limit of coincidence. Physically, this means that the detector is interacting with all frequencies of the field, as the change in the interaction happens in an arbitrarily short time. This is a characteristic feature of  Quantum Field Theory, which would not appear if the field had a finite number of degrees of freedom. 

It is well known, however, that for any state ${\hat\rho_{\textsc{t}}}$ which is Hadamard \cite{wald1994quantum} we can write
\begin{align}
    \braket{\hat{\phi}(\mathsf{x})\hat{\phi}(\mathsf{y})}_{{\hat\rho_{\textsc{t}}}}=F_{{\hat\rho_{\textsc{t}}}}(\mathsf{x},\mathsf{y})+\braket{0|\hat{\phi}(\mathsf{x})\hat{\phi}(\mathsf{y})|0},
\end{align}
where $F_{{\hat\rho_{\textsc{t}}}}(\mathsf{x},\mathsf{y})$ is a regular function and $\braket{0|\hat{\phi}(\mathsf{x})\hat{\phi}(\mathsf{y})|0}$ is the Wightman function of the scalar field in the Minkowski vacuum. As $F_{{\hat\rho_{\textsc{t}}}}(\mathsf{x},\mathsf{y})$ is regular, the second term as $\eta\to 0$ is going to dominate. Once clarified the importance of the vacuum Wightman function in more general grounds, we will dedicate the rest of this appendix to analyze the excitation probability of an UDW detector coupled to the Minkowski vacuum through a nascent delta switching.

In those conditions, the Wightman function takes the form (see, for instance, \cite{takagi1986})
\begin{equation}\label{Wightman}
   \braket{0|\hat{\phi}(\mathsf{x})\hat{\phi}(\mathsf{y})|0}= \int_0^{\infty}\frac{\d\omega}{(2\pi)^d}\frac{D_d(\omega,m)}{2\omega}e^{-\ii\omega \Delta(\mathsf{x}-\mathsf{y})}, \end{equation}
where 
\begin{equation*}
\Delta(\mathsf{x}-\mathsf{y})=\sqrt{-(\mathsf{x}-\mathsf{y})^2}\;\text{sgn}(\mathsf{x}-\mathsf{y})^0   
\end{equation*}
is the space-time interval between $\mathsf{x}$ and $\mathsf{y}$, $d$ is the spatial dimension of the theory and $D_d(\omega,m)$ is the density of states of the field for a certain value of the energy $\omega$ in that dimension. The latter can be written in a closed form \cite{takagi1986}:
\begin{equation}\label{density}
    D_d(\omega,m)=\frac{2^{1-d}\pi^{d/2}}{\Gamma(d/2)}\abs{\omega}(\sqrt{\omega^2-m^2})^{d-2}\Theta(\omega-m).
\end{equation}
 We see that the Wightman function diverges for Klein-Gordon fields in the coincidence limit, that is for $\mathsf{x}=\mathsf{y}$, as their density of states is unbounded.

Therefore, we particularize the functional \eqref{functional} to 
\begin{equation}\label{functionalvacuum}
    \mathcal{W}^\Omega_{0, d}[f,g]=\int^{\infty}_{-\infty}\!\!\!\! \d\tau \!\int^{\infty}_{-\infty}\!\!\!\!\!  \d\tau'f(\tau)[g(\tau')]^*  e^{-\ii\Omega (\tau-\tau')}W_{0,d}(\tau,\tau'),
\end{equation}
where
\begin{equation}
W_{0,d}(\tau,\tau')= \braket{0|\hat{\phi}\big(\mathsf{x}(\tau)\big)\hat{\phi}\big(\mathsf{x}(\tau')\big)|0}.
\end{equation}
 Hence, taking into account the formulae \eqref{functionalvacuum} and \eqref{Wightman} and for  a switching function of the form \eqref{nascent}, we have that the functional \eqref{functional} is
\begin{align}\label{functionalzero}
\nn &\mathcal{W}_{0, d}[\varphi_\eta,\varphi_\eta]=\int_0^{\infty}\!\!\!\!\frac{\d\omega}{(2\pi)^d}\frac{D_d(\omega)}{2\omega}\int^{\infty}_{-\infty}\int^{\infty}_{-\infty}\!\!\!\!\! \d\tau \d\tau'\\
  &\times\frac{1}{\eta^2}\varphi(\tau/\eta)\varphi(\tau'/\eta)  e^{-\ii\left(\omega \Delta\big(\mathsf{x}(\tau)-\mathsf{y}(\tau')\big)+\Omega (\tau-\tau')\right)}.
\end{align}
 In order to analyze the asymptotic behavior as $\eta\to0$, we perform the following changes of variables:
\begin{align}
    \bar\omega&=\omega\eta, &\bar\tau&=\tau/\eta, &\bar\tau'&=\tau'/\eta,
\end{align}
so \eqref{functionalzero} takes the form
\begin{align}
  \nn \mathcal{W}^\Omega_{0, d}&[\varphi_\eta,\varphi_\eta]= \int_0^{\infty}\!\!\!\!\frac{\d\bar\omega}{(2\pi)^d}\frac{D_d(\bar\omega/\eta,m)}{2\bar\omega}\int^{\infty}_{-\infty}\int^{\infty}_{-\infty}\!\!\!\!\! \d\bar\tau \d\bar\tau'\\
  &\times\varphi(\bar\tau)\varphi(\bar\tau')  e^{-\ii\left(\bar\omega \Delta\big(\mathsf{x}(\eta\bar\tau)-\mathsf{y}(\eta\bar\tau')\big)/\eta+\Omega\eta (\bar\tau-\bar\tau')\right)}.
\end{align}
Now, it can be seen that the density of states from \eqref{density} fulfills the following:
\begin{align}\label{homodensisty}
    D_d(\bar\omega/\eta,m)=\eta^{1-d}D_d(\bar\omega,m\eta).
\end{align}
For $d\geq1$ we write 
\begin{align}
  \nn \mathcal{W}^\Omega_{0, d}&[\varphi_\eta,\varphi_\eta]= \eta^{1-d}\int_0^{\infty}\!\!\!\!\frac{\d\bar\omega}{(2\pi)^d}\frac{D_d(\bar\omega,m\eta)}{2\bar\omega}\int^{\infty}_{-\infty}\int^{\infty}_{-\infty}\!\!\!\!\! \d\bar\tau \d\bar\tau'\\
  &\times\varphi(\bar\tau)\varphi(\bar\tau')  e^{-\ii\left(\bar\omega \Delta\big(\mathsf{x}(\eta\bar\tau)-\mathsf{y}(\eta\bar\tau')\big)/\eta+\Omega\eta (\bar\tau-\bar\tau')\right)}.
\end{align}
Finally, assuming that the conditions of the dominated convergence theorem \cite{bartle2014elements} are fulfilled in the integrals we can take the limit of the quantity $\eta^{d-1}\mathcal{W}_{0, d}[\varphi_\eta,\varphi_\eta]$ (for $d>1$) when $\eta$ goes to zero inside the integral. This limit leads to the asymptotic behavior of the excitation probability:
\begin{align}
    &\nn \lim_{\eta\to0}\eta^{d-1}\mathcal{W}_{0, d}[\varphi_\eta,\varphi_\eta]\\
    &\nn =\int_0^{\infty}\!\!\!\!\frac{\d\bar\omega}{(2\pi)^d}\frac{D_d(\bar\omega,0)}{2\bar\omega}\int^{\infty}_{-\infty}\int^{\infty}_{-\infty}\!\!\!\!\! \d\bar\tau \d\bar\tau'\varphi(\bar\tau)\varphi(\bar\tau')  e^{-\ii\bar\omega (\tau-\tau')}\\
  & =\int_0^{\infty}\!\!\!\!\frac{\d\bar\omega}{(2\pi)^d}\frac{D_d(\bar\omega,0)}{2\bar\omega}\abs{\tilde\varphi(\bar{\omega})}^2,
\end{align}
where $\tilde{\varphi}$ is the Fourier transform of the function $\varphi$, which we assume that decays faster than any polynomial. Also, we have taken into account that
\begin{align}
\nn\lim_{\eta\to0}&\frac{\Delta(\eta\bar\tau,\eta\bar\tau')}{\eta}\\
\nn&=\lim_{\eta\to0}\frac{1}{\eta}\sqrt{-(\mathsf{x}(\eta\bar\tau)\!-\!\mathsf{x}(\eta\bar\tau'))^2}\;\text{sgn}\left(\eta(\bar\tau\!-\!\bar\tau')\right)\\
&=\abs{\dot{\mathsf{x}}^2(0)}(\bar\tau\!-\!\bar\tau')= \bar\tau\!-\!\bar\tau',    
\end{align}
where we have expanded the numerator at first order in $\eta$ and we have taken into account that any time-like trajectory fulfills $\dot{\mathsf{x}}^2=-1$.
Thus, we can write the leading order behavior of the excitation probability as $\eta\to0$ as
\begin{align}\label{asym1}
P^+\sim\lambda^2 \eta^{1-d} \int_0^{\infty}\!\!\!\!\frac{\d\bar\omega}{(2\pi)^d}\frac{D_d(\bar\omega,0)}{2\bar\omega}\abs{\tilde\varphi(\bar{\omega})}^2,
\end{align}
with $d>1$. The case $d=1$ is special because the divergence is not only ultraviolet, but also infrared, that is for $m=0$ the spectral density $D_1(\omega,0)/2\omega$ is not bounded from below. We expect, however, a logarithmic divergence as $m$ goes to zero (see, for instance, \cite{takagi1986}). We see from equation \eqref{homodensisty}, that the massless limit and our delta switching limit are the same in $1+1$ dimensions.  

We see from \eqref{asym1} that the response of the detector ignores both the mass of the field and the energy gap of the detector for $d>1$. Therefore the detector's response is asymptotically independent on its state of motion as the interaction time goes to zero. We observe that the detector interacts with the field as if it were following an inertial trajectory. Note that as the excitation probability has to be Lorentz invariant \cite{Martin-Martinez2018}, it does not depend on the velocity of the detector.

We have studied the structure of the divergences of an UDW detector when it is coupled to a Klein-Gordon field for arbitrary mass and arbitrary dimensions, and for arbitrary Hadamard states. Actually this analysis is auxiliary to the main purpose of this paper as we remarked at the beginning of the subsection; we are not interested in concrete expressions for individual excitation probabilities. However, what we learn from this study is that the individual excitation probabilities are not well defined in the exact limit $\eta\to0$, so we can only provide asymptotic expressions.  In our opinion, this does not present a problem as any experimental device will introduce a certain amount of error. The important conclusion is that we can probe the Wightman function as accurately as we want by shortening the duration of the interactions $\eta$.

\bibliography{refs}

\begin{thebibliography}{25}%
\makeatletter
\providecommand \@ifxundefined [1]{%
 \@ifx{#1\undefined}
}%
\providecommand \@ifnum [1]{%
 \ifnum #1\expandafter \@firstoftwo
 \else \expandafter \@secondoftwo
 \fi
}%
\providecommand \@ifx [1]{%
 \ifx #1\expandafter \@firstoftwo
 \else \expandafter \@secondoftwo
 \fi
}%
\providecommand \natexlab [1]{#1}%
\providecommand \enquote  [1]{``#1''}%
\providecommand \bibnamefont  [1]{#1}%
\providecommand \bibfnamefont [1]{#1}%
\providecommand \citenamefont [1]{#1}%
\providecommand \href@noop [0]{\@secondoftwo}%
\providecommand \href [0]{\begingroup \@sanitize@url \@href}%
\providecommand \@href[1]{\@@startlink{#1}\@@href}%
\providecommand \@@href[1]{\endgroup#1\@@endlink}%
\providecommand \@sanitize@url [0]{\catcode `\\12\catcode `\$12\catcode
  `\&12\catcode `\#12\catcode `\^12\catcode `\_12\catcode `\%12\relax}%
\providecommand \@@startlink[1]{}%
\providecommand \@@endlink[0]{}%
\providecommand \url  [0]{\begingroup\@sanitize@url \@url }%
\providecommand \@url [1]{\endgroup\@href {#1}{\urlprefix }}%
\providecommand \urlprefix  [0]{URL }%
\providecommand \Eprint [0]{\href }%
\providecommand \doibase [0]{http://dx.doi.org/}%
\providecommand \selectlanguage [0]{\@gobble}%
\providecommand \bibinfo  [0]{\@secondoftwo}%
\providecommand \bibfield  [0]{\@secondoftwo}%
\providecommand \translation [1]{[#1]}%
\providecommand \BibitemOpen [0]{}%
\providecommand \bibitemStop [0]{}%
\providecommand \bibitemNoStop [0]{.\EOS\space}%
\providecommand \EOS [0]{\spacefactor3000\relax}%
\providecommand \BibitemShut  [1]{\csname bibitem#1\endcsname}%
\let\auto@bib@innerbib\@empty
\bibitem [{\citenamefont {Mart\'{\i}n-Mart\'{\i}nez}\ and\ \citenamefont
  {Rodriguez-Lopez}(2018)}]{Martin-Martinez2018}%
  \BibitemOpen
  \bibfield  {author} {\bibinfo {author} {\bibfnamefont {E.}~\bibnamefont
  {Mart\'{\i}n-Mart\'{\i}nez}}\ and\ \bibinfo {author} {\bibfnamefont
  {P.}~\bibnamefont {Rodriguez-Lopez}},\ }\href {\doibase
  10.1103/PhysRevD.97.105026} {\bibfield  {journal} {\bibinfo  {journal} {Phys.
  Rev. D}\ }\textbf {\bibinfo {volume} {97}},\ \bibinfo {pages} {105026}
  (\bibinfo {year} {2018})}\BibitemShut {NoStop}%
\bibitem [{\citenamefont {Unruh}(1976)}]{Unruh:1976aa}%
  \BibitemOpen
  \bibfield  {author} {\bibinfo {author} {\bibfnamefont {W.~G.}\ \bibnamefont
  {Unruh}},\ }\href {\doibase 10.1103/PhysRevD.14.870} {\bibfield  {journal}
  {\bibinfo  {journal} {Phys. Rev. D}\ }\textbf {\bibinfo {volume} {14}},\
  \bibinfo {pages} {870} (\bibinfo {year} {1976})}\BibitemShut {NoStop}%
\bibitem [{\citenamefont {Hawking}(1975)}]{Hawking1975}%
  \BibitemOpen
  \bibfield  {author} {\bibinfo {author} {\bibfnamefont {S.~W.}\ \bibnamefont
  {Hawking}},\ }\href {\doibase 10.1007/BF02345020} {\bibfield  {journal}
  {\bibinfo  {journal} {Comm. Math. Phys.}\ }\textbf {\bibinfo {volume} {43}},\
  \bibinfo {pages} {199} (\bibinfo {year} {1975})}\BibitemShut {NoStop}%
\bibitem [{\citenamefont {Fewster}\ \emph {et~al.}(2016)\citenamefont
  {Fewster}, \citenamefont {Ju{\'a}rez-Aubry},\ and\ \citenamefont
  {Louko}}]{fewsterjorma}%
  \BibitemOpen
  \bibfield  {author} {\bibinfo {author} {\bibfnamefont {C.~J.}\ \bibnamefont
  {Fewster}}, \bibinfo {author} {\bibfnamefont {B.~A.}\ \bibnamefont
  {Ju{\'a}rez-Aubry}}, \ and\ \bibinfo {author} {\bibfnamefont
  {J.}~\bibnamefont {Louko}},\ }\href
  {http://stacks.iop.org/0264-9381/33/i=16/a=165003} {\bibfield  {journal}
  {\bibinfo  {journal} {Class. Quantum Grav.}\ }\textbf {\bibinfo {volume}
  {33}},\ \bibinfo {pages} {165003} (\bibinfo {year} {2016})}\BibitemShut
  {NoStop}%
\bibitem [{\citenamefont {Garay}\ \emph {et~al.}(2016)\citenamefont {Garay},
  \citenamefont {Mart\'{\i}n-Mart\'{\i}nez},\ and\ \citenamefont
  {de~Ram\'on}}]{Antiunruh}%
  \BibitemOpen
  \bibfield  {author} {\bibinfo {author} {\bibfnamefont {L.~J.}\ \bibnamefont
  {Garay}}, \bibinfo {author} {\bibfnamefont {E.}~\bibnamefont
  {Mart\'{\i}n-Mart\'{\i}nez}}, \ and\ \bibinfo {author} {\bibfnamefont
  {J.}~\bibnamefont {de~Ram\'on}},\ }\href {\doibase
  10.1103/PhysRevD.94.104048} {\bibfield  {journal} {\bibinfo  {journal} {Phys.
  Rev. D}\ }\textbf {\bibinfo {volume} {94}},\ \bibinfo {pages} {104048}
  (\bibinfo {year} {2016})}\BibitemShut {NoStop}%
\bibitem [{\citenamefont {Louko}\ and\ \citenamefont
  {Satz}(2006)}]{louko2006often}%
  \BibitemOpen
  \bibfield  {author} {\bibinfo {author} {\bibfnamefont {J.}~\bibnamefont
  {Louko}}\ and\ \bibinfo {author} {\bibfnamefont {A.}~\bibnamefont {Satz}},\
  }\href@noop {} {\bibfield  {journal} {\bibinfo  {journal} {Class. Quantum
  Grav.}\ }\textbf {\bibinfo {volume} {23}},\ \bibinfo {pages} {6321} (\bibinfo
  {year} {2006})}\BibitemShut {NoStop}%
\bibitem [{\citenamefont {Hotta}(2008)}]{Hotta:2008aa}%
  \BibitemOpen
  \bibfield  {author} {\bibinfo {author} {\bibfnamefont {M.}~\bibnamefont
  {Hotta}},\ }\href {\doibase 10.1103/PhysRevD.78.045006} {\bibfield  {journal}
  {\bibinfo  {journal} {Phys. Rev. D}\ }\textbf {\bibinfo {volume} {78}},\
  \bibinfo {pages} {045006} (\bibinfo {year} {2008})}\BibitemShut {NoStop}%
\bibitem [{\citenamefont {Pozas-Kerstjens}\ \emph {et~al.}(2017)\citenamefont
  {Pozas-Kerstjens}, \citenamefont {Louko},\ and\ \citenamefont
  {Mart\'{\i}n-Mart\'{\i}nez}}]{PozasJorma}%
  \BibitemOpen
  \bibfield  {author} {\bibinfo {author} {\bibfnamefont {A.}~\bibnamefont
  {Pozas-Kerstjens}}, \bibinfo {author} {\bibfnamefont {J.}~\bibnamefont
  {Louko}}, \ and\ \bibinfo {author} {\bibfnamefont {E.}~\bibnamefont
  {Mart\'{\i}n-Mart\'{\i}nez}},\ }\href {\doibase 10.1103/PhysRevD.95.105009}
  {\bibfield  {journal} {\bibinfo  {journal} {Phys. Rev. D}\ }\textbf {\bibinfo
  {volume} {95}},\ \bibinfo {pages} {105009} (\bibinfo {year}
  {2017})}\BibitemShut {NoStop}%
\bibitem [{\citenamefont {Simidzija}\ and\ \citenamefont
  {Mart\'{\i}n-Mart\'{\i}nez}(2017{\natexlab{a}})}]{Petarpetandolo}%
  \BibitemOpen
  \bibfield  {author} {\bibinfo {author} {\bibfnamefont {P.}~\bibnamefont
  {Simidzija}}\ and\ \bibinfo {author} {\bibfnamefont {E.}~\bibnamefont
  {Mart\'{\i}n-Mart\'{\i}nez}},\ }\href {\doibase 10.1103/PhysRevD.96.065008}
  {\bibfield  {journal} {\bibinfo  {journal} {Phys. Rev. D}\ }\textbf {\bibinfo
  {volume} {96}},\ \bibinfo {pages} {065008} (\bibinfo {year}
  {2017}{\natexlab{a}})}\BibitemShut {NoStop}%
\bibitem [{\citenamefont {Simidzija}\ and\ \citenamefont
  {Mart\'{\i}n-Mart\'{\i}nez}(2017{\natexlab{b}})}]{Petarpetandolo1}%
  \BibitemOpen
  \bibfield  {author} {\bibinfo {author} {\bibfnamefont {P.}~\bibnamefont
  {Simidzija}}\ and\ \bibinfo {author} {\bibfnamefont {E.}~\bibnamefont
  {Mart\'{\i}n-Mart\'{\i}nez}},\ }\href {\doibase 10.1103/PhysRevD.96.025020}
  {\bibfield  {journal} {\bibinfo  {journal} {Phys. Rev. D}\ }\textbf {\bibinfo
  {volume} {96}},\ \bibinfo {pages} {025020} (\bibinfo {year}
  {2017}{\natexlab{b}})}\BibitemShut {NoStop}%
\bibitem [{\citenamefont {Lin}\ and\ \citenamefont {Hu}(2007)}]{GrandBei}%
  \BibitemOpen
  \bibfield  {author} {\bibinfo {author} {\bibfnamefont {S.-Y.}\ \bibnamefont
  {Lin}}\ and\ \bibinfo {author} {\bibfnamefont {B.~L.}\ \bibnamefont {Hu}},\
  }\href {\doibase 10.1103/PhysRevD.76.064008} {\bibfield  {journal} {\bibinfo
  {journal} {Phys. Rev. D}\ }\textbf {\bibinfo {volume} {76}},\ \bibinfo
  {pages} {064008} (\bibinfo {year} {2007})}\BibitemShut {NoStop}%
\bibitem [{\citenamefont {Hu}\ \emph {et~al.}(2012)\citenamefont {Hu},
  \citenamefont {Lin},\ and\ \citenamefont {Louko}}]{Beilok}%
  \BibitemOpen
  \bibfield  {author} {\bibinfo {author} {\bibfnamefont {B.~L.}\ \bibnamefont
  {Hu}}, \bibinfo {author} {\bibfnamefont {S.-Y.}\ \bibnamefont {Lin}}, \ and\
  \bibinfo {author} {\bibfnamefont {J.}~\bibnamefont {Louko}},\ }\href
  {http://stacks.iop.org/0264-9381/29/i=22/a=224005} {\bibfield  {journal}
  {\bibinfo  {journal} {Class. Quantum Grav.}\ }\textbf {\bibinfo {volume}
  {29}},\ \bibinfo {pages} {224005} (\bibinfo {year} {2012})}\BibitemShut
  {NoStop}%
\bibitem [{\citenamefont {Bruschi}\ \emph {et~al.}(2013)\citenamefont
  {Bruschi}, \citenamefont {Lee},\ and\ \citenamefont {Fuentes}}]{Fuentas}%
  \BibitemOpen
  \bibfield  {author} {\bibinfo {author} {\bibfnamefont {D.~E.}\ \bibnamefont
  {Bruschi}}, \bibinfo {author} {\bibfnamefont {A.~R.}\ \bibnamefont {Lee}}, \
  and\ \bibinfo {author} {\bibfnamefont {I.}~\bibnamefont {Fuentes}},\ }\href
  {http://stacks.iop.org/1751-8121/46/i=16/a=165303} {\bibfield  {journal}
  {\bibinfo  {journal} {J. of Phys. A: Math. Theor.}\ }\textbf {\bibinfo
  {volume} {46}},\ \bibinfo {pages} {165303} (\bibinfo {year}
  {2013})}\BibitemShut {NoStop}%
\bibitem [{\citenamefont {Brown}\ \emph {et~al.}(2013)\citenamefont {Brown},
  \citenamefont {Mart\'{\i}n-Mart\'{\i}nez}, \citenamefont {Menicucci},\ and\
  \citenamefont {Mann}}]{Eric}%
  \BibitemOpen
  \bibfield  {author} {\bibinfo {author} {\bibfnamefont {E.~G.}\ \bibnamefont
  {Brown}}, \bibinfo {author} {\bibfnamefont {E.}~\bibnamefont
  {Mart\'{\i}n-Mart\'{\i}nez}}, \bibinfo {author} {\bibfnamefont {N.~C.}\
  \bibnamefont {Menicucci}}, \ and\ \bibinfo {author} {\bibfnamefont {R.~B.}\
  \bibnamefont {Mann}},\ }\href {\doibase 10.1103/PhysRevD.87.084062}
  {\bibfield  {journal} {\bibinfo  {journal} {Phys. Rev. D}\ }\textbf {\bibinfo
  {volume} {87}},\ \bibinfo {pages} {084062} (\bibinfo {year}
  {2013})}\BibitemShut {NoStop}%
\bibitem [{\citenamefont {Simidzija}\ \emph {et~al.}(2018)\citenamefont
  {Simidzija}, \citenamefont {Jonsson},\ and\ \citenamefont
  {Mart\'{\i}n-Mart\'{\i}nez}}]{Petarpetandolo3}%
  \BibitemOpen
  \bibfield  {author} {\bibinfo {author} {\bibfnamefont {P.}~\bibnamefont
  {Simidzija}}, \bibinfo {author} {\bibfnamefont {R.~H.}\ \bibnamefont
  {Jonsson}}, \ and\ \bibinfo {author} {\bibfnamefont {E.}~\bibnamefont
  {Mart\'{\i}n-Mart\'{\i}nez}},\ }\href {\doibase 10.1103/PhysRevD.97.125002}
  {\bibfield  {journal} {\bibinfo  {journal} {Phys. Rev. D}\ }\textbf {\bibinfo
  {volume} {97}},\ \bibinfo {pages} {125002} (\bibinfo {year}
  {2018})}\BibitemShut {NoStop}%
\bibitem [{\citenamefont {Bartle}(1995)}]{bartle2014elements}%
  \BibitemOpen
  \bibfield  {author} {\bibinfo {author} {\bibfnamefont {R.~G.}\ \bibnamefont
  {Bartle}},\ }\href@noop {} {\emph {\bibinfo {title} {The Elements of
  Integration and Lebesgue Measure}}}\ (\bibinfo  {publisher} {John Wiley \&
  Sons},\ \bibinfo {address} {New York},\ \bibinfo {year} {1995})\BibitemShut
  {NoStop}%
\bibitem [{\citenamefont {Berberian}\ \emph {et~al.}(1974)\citenamefont
  {Berberian} \emph {et~al.}}]{berberian1974lectures}%
  \BibitemOpen
  \bibfield  {author} {\bibinfo {author} {\bibfnamefont {S.~K.}\ \bibnamefont
  {Berberian}} \emph {et~al.},\ }\href@noop {} {\emph {\bibinfo {title}
  {Lectures in functional analysis and operator theory}}},\ Vol.~\bibinfo
  {volume} {15}\ (\bibinfo  {publisher} {Springer New York, Heidelberg,
  Berlin},\ \bibinfo {year} {1974})\BibitemShut {NoStop}%
\bibitem [{\citenamefont {Wald}(1994)}]{wald1994quantum}%
  \BibitemOpen
  \bibfield  {author} {\bibinfo {author} {\bibfnamefont {R.~M.}\ \bibnamefont
  {Wald}},\ }\href@noop {} {\emph {\bibinfo {title} {Quantum field theory in
  curved spacetime and black hole thermodynamics}}}\ (\bibinfo  {publisher}
  {University of Chicago Press},\ \bibinfo {year} {1994})\BibitemShut {NoStop}%
\bibitem [{\citenamefont {Pozas-Kerstjens}\ and\ \citenamefont
  {Mart{\'\i}n-Mart{\'\i}nez}(2016)}]{pozas2016entanglement}%
  \BibitemOpen
  \bibfield  {author} {\bibinfo {author} {\bibfnamefont {A.}~\bibnamefont
  {Pozas-Kerstjens}}\ and\ \bibinfo {author} {\bibfnamefont {E.}~\bibnamefont
  {Mart{\'\i}n-Mart{\'\i}nez}},\ }\href@noop {} {\bibfield  {journal} {\bibinfo
   {journal} {Phys. Rev. D}\ }\textbf {\bibinfo {volume} {94}},\ \bibinfo
  {pages} {064074} (\bibinfo {year} {2016})}\BibitemShut {NoStop}%
\bibitem [{\citenamefont {DeWitt}(1979)}]{dewitt1979quantum}%
  \BibitemOpen
  \bibfield  {author} {\bibinfo {author} {\bibfnamefont {B.}~\bibnamefont
  {DeWitt}},\ }\href@noop {} {\emph {\bibinfo {title} {Quantum gravity: the new
  synthesis in General Relativity: An Einstein centenary survey, Eds. SW
  Hawking and W. Israel}}}\ (\bibinfo  {publisher} {Cambridge University Press,
  Cambridge},\ \bibinfo {year} {1979})\BibitemShut {NoStop}%
\bibitem [{\citenamefont {Birrell}\ and\ \citenamefont
  {Davies}(1984)}]{birrell1984quantum}%
  \BibitemOpen
  \bibfield  {author} {\bibinfo {author} {\bibfnamefont {N.~D.}\ \bibnamefont
  {Birrell}}\ and\ \bibinfo {author} {\bibfnamefont {P.}~\bibnamefont
  {Davies}},\ }\href@noop {} {\emph {\bibinfo {title} {Quantum fields in curved
  space}}}\ (\bibinfo  {publisher} {Cambridge University Press},\ \bibinfo
  {year} {1984})\BibitemShut {NoStop}%
\bibitem [{\citenamefont {Takagi}(1986)}]{takagi1986}%
  \BibitemOpen
  \bibfield  {author} {\bibinfo {author} {\bibfnamefont {S.}~\bibnamefont
  {Takagi}},\ }\href {\doibase 10.1143/PTP.88.1} {\bibfield  {journal}
  {\bibinfo  {journal} {Prog. Theor. Phys. Suppl.}\ }\textbf {\bibinfo {volume}
  {88}},\ \bibinfo {pages} {1} (\bibinfo {year} {1986})}\BibitemShut {NoStop}%
\bibitem [{\citenamefont {De~Bi{\`e}vre}\ and\ \citenamefont
  {Merkli}(2006)}]{bievre2006}%
  \BibitemOpen
  \bibfield  {author} {\bibinfo {author} {\bibfnamefont {S.}~\bibnamefont
  {De~Bi{\`e}vre}}\ and\ \bibinfo {author} {\bibfnamefont {M.}~\bibnamefont
  {Merkli}},\ }\href@noop {} {\bibfield  {journal} {\bibinfo  {journal} {Class.
  Quantum Grav.}\ }\textbf {\bibinfo {volume} {23}},\ \bibinfo {pages} {6525}
  (\bibinfo {year} {2006})}\BibitemShut {NoStop}%
\bibitem [{\citenamefont {Carballo-Rubio}\ \emph {et~al.}(2018)\citenamefont
  {Carballo-Rubio}, \citenamefont {Garay}, \citenamefont {Martin-Martinez},\
  and\ \citenamefont {de~Ramon}}]{Carballo-Rubio:2018aa}%
  \BibitemOpen
  \bibfield  {author} {\bibinfo {author} {\bibfnamefont {R.}~\bibnamefont
  {Carballo-Rubio}}, \bibinfo {author} {\bibfnamefont {L.~J.}\ \bibnamefont
  {Garay}}, \bibinfo {author} {\bibfnamefont {E.}~\bibnamefont
  {Martin-Martinez}}, \ and\ \bibinfo {author} {\bibfnamefont {J.}~\bibnamefont
  {de~Ramon}},\ }\href {https://arxiv.org/abs/1804.00685} {\  (\bibinfo {year}
  {2018})},\ \Eprint {http://arxiv.org/abs/1804.00685} {1804.00685}
  \BibitemShut {NoStop}%
\bibitem [{\citenamefont {Schlicht}(2004)}]{schlicht2004considerations}%
  \BibitemOpen
  \bibfield  {author} {\bibinfo {author} {\bibfnamefont {S.}~\bibnamefont
  {Schlicht}},\ }\href@noop {} {\bibfield  {journal} {\bibinfo  {journal}
  {Class. Quantum Grav.}\ }\textbf {\bibinfo {volume} {21}},\ \bibinfo {pages}
  {4647} (\bibinfo {year} {2004})}\BibitemShut {NoStop}%
\end{thebibliography}%

\end{document}